\documentclass[useAMS,usenatbib]{mn2e}
\usepackage{graphics}
\usepackage{subfigure}
\usepackage{graphicx}
\usepackage{amssymb}

\title[Automated detection of Filaments in the LSS]{Automated detection of filaments in the large scale structure of the Universe}

\author[Roberto E. Gonz\'alez and Nelson D. Padilla]{Roberto E. Gonz\'alez$^{1}$\thanks{E-mail:
regonzar@astro.puc.cl (REG); npadilla@astro.puc.cl (NDP)} and Nelson D. Padilla$^{1}$\\
$^{1}$Departamento de Astronom\'\i{}a y Astrof\'\i{}sica, Pontificia Universidad Cat\'olica de Chile, Santiago, Chile\\
}

\begin{document}

\date{Accepted --- . Received ---}

\pagerange{\pageref{firstpage}--\pageref{lastpage}} \pubyear{2009}

\maketitle

\label{firstpage}

\begin{abstract}
We present a new method to identify large scale filaments and apply it to a cosmological simulation.  Using positions of haloes above a given mass as node tracers, we look for filaments between them using the positions and masses of all the remaining dark-matter haloes.  In order to detect a filament, the first step consists in the construction of a backbone linking two nodes, which is given by a skeleton-like path connecting the highest local dark matter (DM) density traced by non-node haloes.
The filament quality is defined by a density and gap parameters characterising its skeleton, and filament members are selected by their binding energy in the plane perpendicular to the filament.  
This membership condition is associated to characteristic orbital times; however if one assumes a fixed orbital timescale for all the filaments,
 {the resulting filament properties show only marginal changes,}
  {indicating that the use of dynamical information is not critical for the method.}
We test the method in the simulation using massive haloes($M>10^{14}$h$^{-1}M_{\odot}$) as filament nodes. 
 {The main properties of the resulting high-quality filaments (which corresponds
to  $\simeq33\%$ of the detected filaments) are,}
i) their lengths cover a wide range of values of up to $150 $h$^{-1}$Mpc, but are mostly concentrated below $50$h$^{-1}$Mpc; ii) their distribution of thickness peaks at $d=3.0$h$^{-1}$Mpc and increases slightly with the filament length; iii) their nodes are connected on average to $1.87\pm0.18$ filaments for $\simeq 10^{14.1}M_{\odot}$ nodes; this number increases with the node mass to $\simeq 2.49\pm0.28$ filaments for $\simeq 10^{14.9}M_{\odot}$ nodes; iv) on average, the central density along the filaments starts at almost a hundred times the average density in the regions surrounding the nodes and then drops to about a few times the mean density at larger distances, where it remains roughly constant over $20$ to $80\%$ of the filament length (this result may depend on the filament length); v) there is a strong relation between length, quality and how straight a filament is, where shorter filaments are those characterised by higher qualities and more straight-line like geometries.
\end{abstract}


\section{Introduction}

The large scale distribution of galaxies and dark matter(DM) shows a web-like structure composed by clusters, 
walls, filaments and void regions, and is usually referred to as the cosmic web. These structures can be 
easily detected by eye in numerical DM simulations or in the observed distribution of galaxies in large surveys 
such as the Sloan Digital Sky Survey \citep[SDSS]{york}.

For clusters and voids, there are several well established automated identification methods which have 
been broadly used, such as the Friend-of-Friends algorithm for halo/cluster detection \citep[FOF]{davis}, 
and the \citet{pa05} algorithm for reliable detection of voids 
 {(see \citealt{col08} for a complete review on different void detection methods).}
In the case of filaments and walls this task is markedly difficult since, in general, there is still no clear consensus on how to characterise them; filaments and walls show complex 3D shapes.

There are different approaches to the study of filaments. From the theoretical point of view it was found  {that the gravitational collapse of matter on large scales leads to the formation of sheets and filaments \citep{zel70}. 
\citet{bon} studied tidal fields in the large-scale structure (LSS) and showed how these produce filamentary structures.} 

There are several sets of filaments which have been identified and characterised by eye in both
simulations and observations. 
{\citet{colb} identified by eye $228$ filaments between massive neighbouring haloes in a DM simulation, and described several interesting statistical properties using this sample.    }
In observations, \citet{pimbblet} and \citet{porter} identified filaments in large surveys by eye, and dark matter (DM) filaments were also be detected between clusters of galaxies using weak lensing techniques \citep{mead}. {In x-ray observations, it has also been possible to detect hot gas filaments connecting clusters \citep{shar}.}

{The study of statistics and the topology of the galaxy distribution with the aim to search for filaments starts very early, with studies by \citet{zel}, \citet{sha}, and \citet{ein}.}
Options to automate the search of filaments include the use of statistics on the morphology of structures, such as Minkowski functionals, minimal spanning trees (MST), percolation methods and shapefinders \citep[see review by][]{martin}. 
The minimum spanning tree method was introduced in cosmology by \citet{barrow}.  This produces a unique graph which connects points of a process without closed loops, but describes mainly the local nearest-neighbour distribution and is unable to provide a full characterisation of the LSS. 
Shapefinders \citep{sahni} have also been used to identify filaments. 

In three dimensions, the morphology of a compact manifold can be characterised by four Minkowski functionals: volume, surface area, integrated mean curvature and integrated gaussian curvature.  It is possible to define a number of quantities related to those functionals; if a set of positions of galaxies or haloes is characterised 
by particular values of ratios between the Minkowski functionals, it is very likely that it will show a filamentary shape \citep{bhar}, but this does not guarantee a true detection of a filament or that all the selected members actually belong to the filament. 

Another algorithm for the detection of filaments was proposed by \citet{pim2} based on the assumption that the orientations of constituent galaxies along such filaments are non-isotropic. This method works well on straight filaments with separations smaller than $15$Mpc$/$h, as has been shown in their application to the 2-degree Field Galaxy Redshift Survey \citep[2dFGRS,][]{col}. 

The Skeleton method \citep{erik,novikov} has proven useful for the detection of possible filamentary structures in continuous two dimensional density fields.  The skeleton is determined by segments parallel to the gradient of the field connecting saddle points to local maxima.  The method involves interpolation and smoothing of the point distribution, introducing the kernel band-width as an extra parameter in the procedure of estimating the density field. 
Extending this work to three dimensions, \citet{sou} found good agreement between detected skeletons and eye detections in a numerical DM simulation.  By using the Hessian matrix eigenvalues they were able to detect filamentary structures \citep[See also][2007b]{arag}.
\citet{bond} also use the Hessian matrix of the galaxy density field smoothed on different scales to characterise the morphology of the LSS in mock catalogues and in the SDSS \citep[][]{stou}; they 
use their detected structures to determine the typical scales where filaments, clumps and walls are dominant. 

The Candy Model used by \citet{stoica1}, is a 
two-dimensional 
marked point process where segments serve as marks.  This method has been adapted to three dimensions and also improved to a more general Bisous Model \citep{stoica2}, producing detections in very good agreement with the result of eye detection in tracing filamentary structures using only galaxy positions (as in the method we will present).  However, the detection and thickness of the resulting filaments is only given by a coverage threshold (percent of total points, to be included in filaments). 

The spin and orientation of haloes in filaments has been studied by \citet{arag2} and \citet{zhang}. They use a Multi-scale Morphology Filter (MMF) and compute the Hessian Matrix eigenvalues in a density field smoothed on different scales, to divide the full volume of their samples into cluster, filament and wall like structures.  However, this method, as well as other Hessian matrix based methods, is affected by a lack of an ability to determine the thickness of filaments, and are difficult to apply to observational data, where one needs to define whether a galaxy is a member of a cluster, filament or void.

In this work we propose a new automated method to detect filaments which builds upon ideas of several of the methods mentioned previously.  A novel feature of the method is that it is designed to search for filaments using nodes (corresponding to haloes or galaxy clusters as in \citealt{colb}) selected by applying lower limits on their mass (or proxy for mass).
This new method aims to be applicable to discrete halo or galaxy positions even when these are so sparsely distributed that it is not possible to define a smooth density field, or that the Hessian matrix cannot be computed with an adequately high accuracy.  This makes it particularly suitable for observational data such as the 2dFGRS or SDSS.
In addition, we replace the smoothing scales and filament coverage thresholds by parameters with improved physical meaning.  In this new approach a filament quality depends on parameters related to the relative density and gaps of the filament skeleton, and its members are identified as the haloes or galaxies with binding energies with respect to the filament in the plane perpendicular to its skeleton.  
We will use the numerical simulations to calibrate the binding condition using objects with a collapse time and radius that can be computed even when dynamical information is not available,  { as is usually the case with observational data.  In the latter, measurements or proxies for galaxy masses will still be required in order to define the filament membership condition.}

This paper is organised as follows.  Section 2 presents the numerical simulation on which we perform our
automated search for filaments.  The method is presented in Section 3, which also includes details on 
the measurement of the local density field, and describes the input parameters of the algorithm.  Section
4 presents the results and Section 5 concludes this work with our conclusions.

\section{The numerical simulation}

\begin{figure*} 
\centering 
\includegraphics[height=.70\linewidth,angle=0]{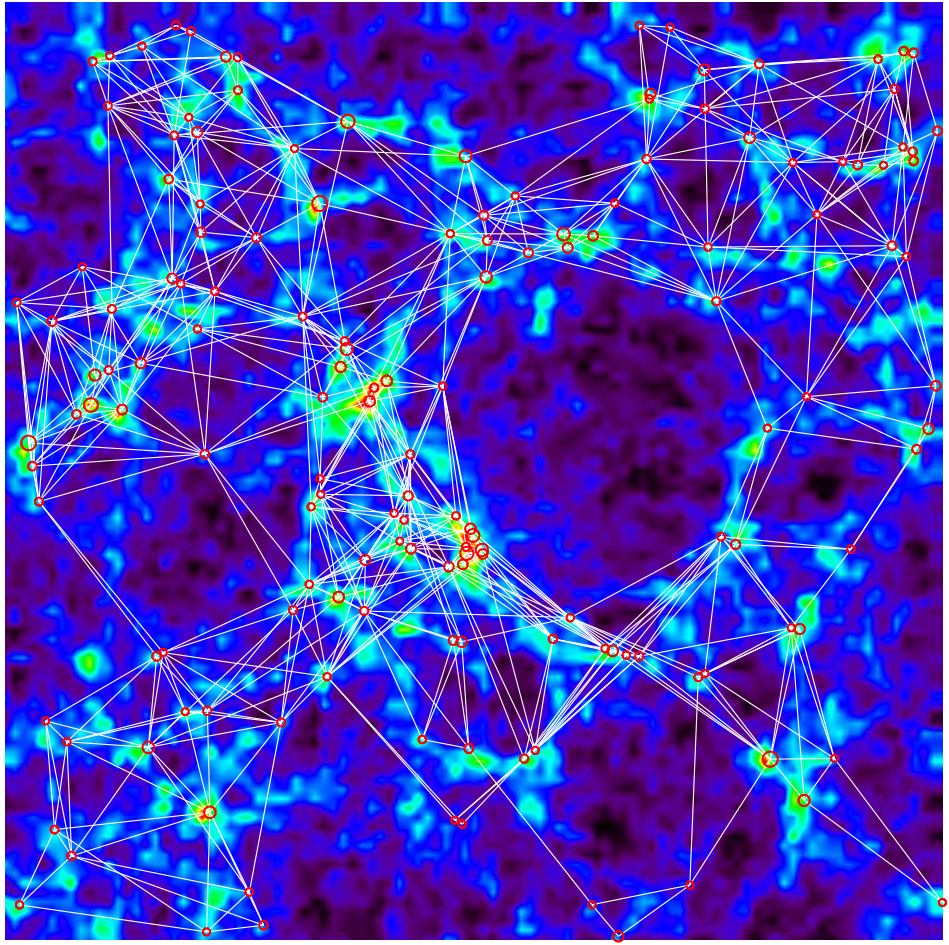}
\caption{
\label{figure1} Density field in the numerical simulation corresponding to a slice $100$h$^{-1}$Mpc thick.  The density is obtained using the halo positions. The red circles enclose the virial radii of the node haloes; white lines connect all the node pairs separated by less than $65$h$^{-1}$Mpc. }

\end{figure*}

We use a cosmological DM simulation with parameters corresponding to the concordance $\Lambda CDM$ model (cold dark-matter, $\Omega_{b}=0.045$, $\Omega_{DM}=0.235$, $\Omega_{DE}=0.72$, $h=0.72$, $\sigma_8=0.847$, \& $n=1$)
, $500^3$ particles and a periodic cube side of $250$Mpc$/$h.  At $z=0$ we find
$176,041$ haloes and subhaloes in the mass range $1.4 \times 10^{11}$h$^{-1}M_{\odot} < M <1.5 \times 10^{15}$h$^{-1}M_{\odot}$, identified using the AHF code \citep{kno}.
For the detection of filaments, we select as nodes a total of $427$ haloes with $M>10^{14}$h$^{-1}M_{\odot}$. The node pairs that will be the candidates for filament search are constructed using neighbour nodes, which are easily obtained using Voronoi Tessellations(VT hereafter, to be explained in more detail in the next section).
We obtain a total of $3,385$ node pairs with separations $<65$h$^{-1}$Mpc,
 {using periodic conditions ($310$ node pairs straddle the simulation borders);} 
Figure \ref{figure1} shows all the 
node pairs in a slice of the simulation. 
In the next section we will apply the filament detection method to each of these node pairs. 

\section{Method}
Our filament detection 
method is described in this section.  
 {We apply the method to dark-matter halo positions in the simulation as a first step towards the 
detection of galaxy 
filaments from observational data.  A future extension will also use halo substructure as well as galaxies
from a semi-analytic model so as to mimick real galaxies as closely as possible (as galaxies are thought
to form in the potential well of DM haloes and subhaloes).
When applying our method to semi-analytic
galaxies we will be able to detect the effects of using proxies for the host dark-matter halo masses obtained
from a galaxy catalogue
(e.g. dynamical masses, luminosities in different bands) instead of the measured dark-matter halo masses.
Finally, our method can also be extended to use redshift-space information to assess the effect
of large-scale bulk motions and the small-scale finger-of-god effect on the resulting filaments.}

We will not attempt to find all the filamentary structures in the simulation, only those filament segments generated between haloes above a given mass threshold (node pairs).  Therefore, smaller filaments associated to less massive nodes will be missed if they are not in the path (or part) of the selected nodes.
 
\subsection{Density field}
 {In this paper we distinguish between two different definitions of density; (i) the standard dark matter density traced by the particles in the simulation which we call DM density; and (ii) the density given by the halo positions and their virial masses which we call the halo density. 
It is clear that the halo density contains little information about the mass and structure that lie beyond the virial radii of the haloes, but as we will show it is still an appropriate proxy of the DM density in the simulation.   It is clear that halo positions and their masses (or in the observational case, galaxy positions and luminosities) allow a clear
by-eye detection of filamentary features at large scales (\citealt{colb}).}

 {In general, }the density and density gradient field  { of a distribution of points can be obtained} using VT, in a similar approach to that adopted by \citet{arag2} where they compute the density field using Delaunay Tessellation Field Estimator \citep{sh01}.  In the present study we make use of the neighbour information for
all the haloes
to trace the  {halo} density field as well as to compute a fast proxy for the  {halo} density gradient vector field.
VT also allows us to obtain the immediate neighbours of each halo (or galaxy if applied to observational data).
The Voronoi Tessellation \citep{voronoi1} technique is one of the best adaptive methods to recover a precise density field from a discrete distribution of points, with clear advantage over the method used in 
Smoothed Particle Hidrodynamic or other interpolation based techniques \citep{pess}.
We compute the VT for the halo distribution defining a cellular-like structure, where each halo is associated to a region (or voronoi cell) in which any point inside this region is nearest to that halo than to any other.

 {This voronoi cell defines a volume which used along with the enclosed mass, defines a very precise and adaptive measure of the density of the cell.}
 {In the case of point masses (such as when using the DM particle distribution), one can measure the exact enclosed mass in each voronoi cell, and therefore compute a very accurate DM density field.}
 {Instead, in this work we use the halo positions along with their measured virial masses. The VT computation is done in the same way as for particles, but the halo virial mass does not account for all the enclosed mass in the voronoi cell, it only includes the mass out to the virial radius.  For instance, in low density environments the halo-to-neighbour distance is much larger than the virial radius, and therefore the mass enclosed in the voronoi cell given by the virial mass of its central halo is underestimated.
The opposite occurs in dense environments where the voronoi cell volume of a halo can be even smaller than their virial sphere due to close neighbours; in this case there is an overestimation of the enclosed mass in the voronoi cell.
As this method does not require absolute density values but only the relative highest density path between nodes (mainly given by the collapsed mass)
the use of the halo density would increase the contrast of filaments improving the ability of the method to follow their high relative density path to some degree.}

 {We argue that in the high density end, the halo density over-estimation is not important for our purposes since i) we 
will not consider subhaloes or haloes inside the virial radius of nodes (the most massive haloes), ii) the inter halo 
distance becomes comparable to the virial radius at halo densities much greater than the average density along the
filaments, and therefore only a few haloes considered in our analysis will suffer this overestimation. 
As a result, most of the haloes that will present an overestimated density will be nodes, and the remaining affected
fraction will be located around nodes and in the central sections of the filaments, where their filament membership 
will be ensured, independently of the overestimation of their density.}

In low density regions the voronoi cells of haloes are always much larger than their virial spheres which produces
an underestimation of the density; later in this section 
we will work on diminishing this problem by using an approximation assuming Navarro, Frenk \& White (1997, NFW) 
profiles,  {to define the characteristic DM density between two haloes.}
 
{  Before moving on to the calculation of the characteristic density between haloes, we will analyse 
in more detail the differences between the halo and DM densities.}
For a smooth density field, such as is the case of fields traced by DM particles, the Hessian matrix can be computed with high accuracy to find the filament components easily.  But the process is more complicated in the case of having only the positions of haloes and their virial masses.  This is due to the sparse coverage of haloes, their variable masses, and the loss of information regarding the mass located beyond the virial radii of haloes.  In order to understand the importance of these issues we will look at the relation between average halo to neighbour separation ($D_{IP}$) and its voronoi cell volume.

\begin{figure*} 
\centering 
\includegraphics[height=.50\linewidth,angle=0]{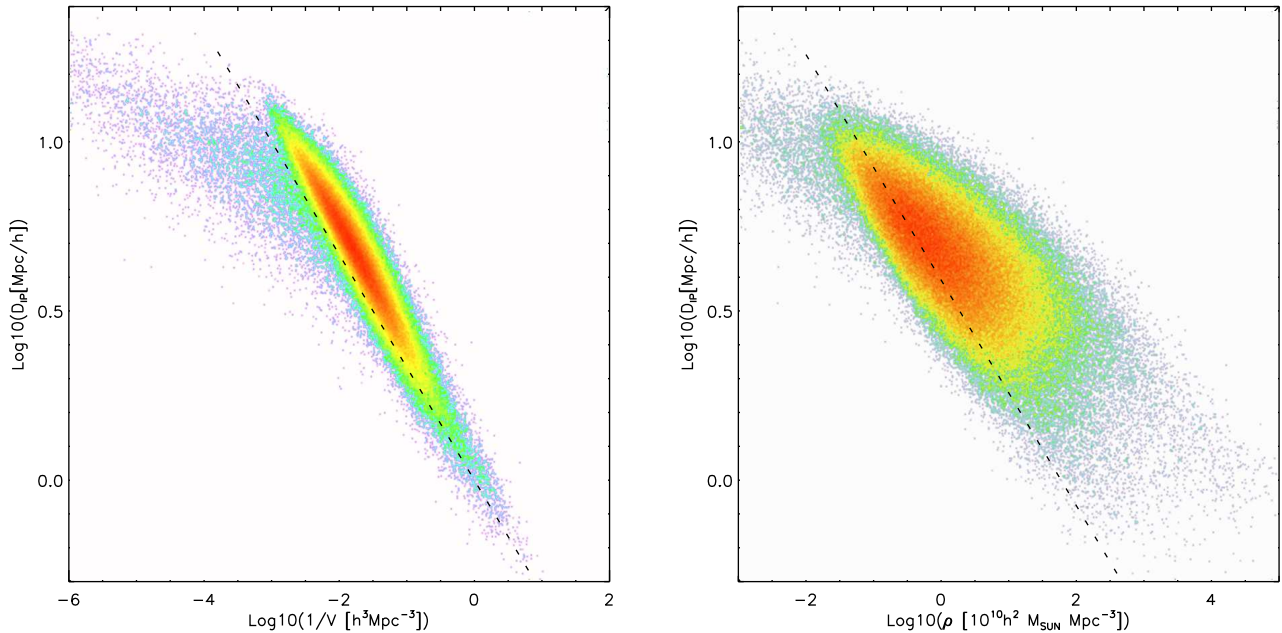}
\caption{
\label{figure2} Left panel: Voronoi cell volume vs. mean neighbour separation, $D_{IP}$, for all the haloes in the simulation. Right panel: Voronoi halo Density vs. $D_{IP}$ for all the haloes. Dashed lines represent the relation $V=D_{IP}^3$ (left panel); in the right panel it corresponds to $V=median(M_{VIR})/\rho$.}
\end{figure*}

In order to recover the real DM density field as best as possible using only halo positions, one needs to take into account that,
\begin{itemize}
\item In high density environments the voronoi cell volume is related to the local mean inter-particle distance, i.e., the mean neighbour distance $D_{IP}$.  The left panel of Figure \ref{figure2} shows a very tight relation between these two quantities for the full halo population.  In the figure, the dashed line shows the $V \propto D_{IP}^3$ relation, which is very useful for  halo detection methods such as FOF \citep{davis}, where the particle separation is used to connect particles above a given density threshold.
In the case of having only halo positions, we find that this relation breaks down at lower densities (as can be seen in the left panel of Figure \ref{figure2}).   The origin of this departure from the distance vs. volume relation is the complex shapes\footnote{We refer as complex cell shapes to non-spherical or non-polyhedric like shapes, produced when having few neighbours at non-uniform distances.} developed by Voronoi cells at such low densities  {as result of greater standard deviation in the computation of $D_{IP}$ due to a low neighbour count and inter-halo distances falling within a wide range of possible values.  Another possibility is that shot noise is affecting our estimates, but
this should not be the main source in our case since haloes mark the highest peaks in the density field, and
we use a relatively large minimum number of particles per halo}. 
This implies that the local clumpiness of a set of  {haloes} in low density environments is only poorly related to its density; this may pose a challenge to the search for the backbone of filaments. This effect is negligible when obtaining the density field using DM particles since these typically produce a smoother spatial coverage and therefore a much smaller fraction of these will be surrounded by Voronoi cells with complex shapes.

\item  {The spread in the virial masses of the haloes, introduces a scatter in the relation 
between mean neighbour separation and the halo density  (right panel of Figure \ref{figure2}) 
with respect to the resulting relation from using only cell volumes. 
Therefore the halo density can only be used as a proxy for the matter density, and will serve to choose
which halo pair will have the highest local DM density; 
by choosing the neighbour with the greatest halo density, it will probably be the nearest one and very likely the correct 
choice.  This will sometimes not be true, for example when two or more neighbours have similar halo densities.  Consider
for example two neighbour cells with almost equal densities, but one having $F$ times more mass and $F$ times more volume than the other($F>1$); 
if we make a simple estimate of the DM density for the region lying between the halo and these two neighbours using NFW 
profiles for each halo, we will find that the path connecting to the smallest and closest neighbour will have the highest 
DM density.  Later in this section we will apply this correction to our VT density estimates.}
\end{itemize}

 {We now estimate the local DM density between a halo and its neighbours, which we call the characteristic DM density $\rho*$.
As we have shown, the halo density estimate is relative and it is only used to find the neighbour with the highest local DM density from all the possible halo-neighbour pairs.  This density is an approximation that depends on the halo masses and inter-halo distances, and therefore it is probably safer not to compare it to the real DM density field given by the DM particles.
Due to these considerations, in order to find the path of highest local DM density connecting two nodes, we need to add conditions on when and how to use of the  {halo} density field.
To estimate the characteristic DM density $\rho*$ between the $i$-th halo and one of its neighbours, halo $j$, we will have two cases depending on the relation between their separation and their virial radii,}

\vskip .3cm
\noindent
i) $D_{ij} \leq R_{VIR}(i)+R_{VIR}(j): \quad \rho*= k_1 \: \rho(j)$, \\
\vskip .3cm
\noindent

\noindent  {where $D_{ij}$ is the distance between haloes, $\rho$ is the halo density, and $k_1$ is a constant which includes the halo density of halo $i$ common to all its neighbours.}
 {The fraction of halo pairs which satisfy this condition is very low and correspond to nodes and their immediate
neighbours (haloes which are linked gravitationally); here the halo density is a good proxy for the DM density, and even a possible over-estimation of 
the halo density due to cell volumes smaller than virial spheres is positive for our purpose, since gravitationally 
linked haloes should have the first priority at the moment of choosing the halo-neighbour path to form the filament skeleton.}
In this case, the segment connecting haloes $i$ and $j$ will have the maximum characteristic DM density among the other immediate neighbours. 

\vskip .3cm
\noindent
ii)$D_{ij} > R_{VIR}(i)+R_{VIR}(j):$ $ \quad \rho*=k_2 \: \rho(j) \: \eta^{-1} f(M_i,M_j)$.\\
\vskip .3cm
\noindent
 {Most halo pairs fall in this second case. Here we}
use NFW profiles to estimate  {a proxy of the characteristic} DM density between two haloes. 
 {This proxy consists on the minimum DM density present in the path between two haloes, obtained by extending 
NFW profiles beyond the halo virial radii (this is a good approximation since the average of the inter-halo separation
in the filament backbones is $4.80\pm0.03$ times the sum of the virial radii of the two neighbour haloes, see Section 4.1).} 
In the equation,
$M_i$ and $M_j$ are the halo masses,
the $\eta$ factor represents the break-down of the relation between inter-halo distance and voronoi cell volume,
\begin{eqnarray}
\eta=\frac{D_{IP}^3}{V_{cell}}, \qquad \qquad \qquad \qquad \qquad \qquad \qquad \qquad \qquad \qquad \nonumber\\
f(M_i,M_j) \simeq \left( \frac{M_i}{M^*} \right)^{0.13} \left( \frac{1+\Omega}{\Omega} \right)^3, \quad \quad \Omega=\left( \frac{M_i}{M_j}\right)^{0.376},
\nonumber
\end{eqnarray}
\noindent
and $k_1$ and $k_2$ are constants intended to provide the continuity between both densities at $\eta=1$, and $\Omega=1$; $M^*=10^{12.5}$h$^{-1}M_{\odot}$ 
is the constant in the \citet{bull} concentration vs. mass relation.
The $\eta$ parameter appears naturally in this approximation where its value is usually greater than one; 
therefore, two haloes with high masses and high voronoi halo densities will have lower $\rho*$ if their separation 
is large, as can be the case in regions with a low number density of discrete points.

The DM density between two haloes will be used as segment weights in the search for the path connecting two nodes, in a similar way to that used in the search for the shortest path in graph theory; therefore, the filament backbone or skeleton is the result of solving for this graph, which has several different approaches in the literature (Biggs,
Lloyd \& Wilson, 1986). 

\subsection{Input parameters}

We detect filaments using nodes above a fixed minimum mass.  This choice is necessary since the filamentary structure is found at different scales; there are even filaments inside filaments or inside clusters \citep{bon}.

In addition to the minimum node mass, other parameters will be necessary
 {since otherwise it is always possible to find the highest density path connecting any two nodes.}
However, our aim is to involve only the lowest number of parameters possible, which include the following,
\begin{itemize}
\item A minimum density threshold for the galaxies or haloes which form the backbone of a filament. 
 {This density refers to a minimum characteristic DM density (defined in the previous subsection) along the consecutive halo pairs which form the filament backbone.}
 {There is no fixed physically motivated minimum value for this quantity, but we are interested in the 
filaments which are at least noticeably above the local background density, i.e. filament backbones above 
a few times the mean density. We will use this minimum density as a quality parameter for the detected filaments, 
since the higher this density for a filament is, the stronger the density gradients and filament-like 
potential will be, with more haloes bounded to them.} 
\item A maximum gap threshold for the galaxies or haloes which define the backbone of the filament.  A measure of the gaps in a filament is given by \mbox{$max(D_{SK}/<D_{SK}>)$}, the maximum distance divided by the average distance between all pairs of consecutive skeleton members of an individual filament.  Large values for this parameter imply large gaps between two filament sections.  Gaps are an important problem, particularly for low density filaments.  {Again, this parameter will not define a limit on what is identified as a filament, but will be used as another quality parameter since the smaller this value is, the more continuous and uniform the filament will be, with less noticeable gaps in the backbone.}
\item After the definition of the backbone or skeleton of the filament has been completed, we select the members of the filament.  This is done by analysing which neighbours are gravitationally linked to the filament and will collapse into the skeleton or remain within the filament for at least a given amount of time. We define a timescale $t_F$, which is the maximum time allowed for the orbit of a halo in the plane perpendicular to the filament, assuming it is gravitationally bound (in this plane). Since the peculiar velocities of the haloes in the numerical simulation are known, we can calculate which haloes are bound to the filament; we use this information to characterise an average timescale and the associated radius out to which bound haloes can be found.  This will help to implement this filament identification in the case of observational data with no available information on peculiar velocities.
\end{itemize}

It is complicated to define physically motivated density and gap thresholds for each filament analogous to the virialisation density for the spherical collapse model.  The reasons behind this are the complicated filament shapes and their continuous feeding of their node haloes or clusters.  
Therefore, we will use these parameters to assess the quality of a filament; filaments will be better defined if their minimum backbone densities are high and their largest gaps are small.  

 {The reasons behind the choice of these two parameters to define the quality of filaments are the following. 
A filament is a region in the universe where the gravitational collapse of matter occurs mainly towards a line 
(continous but not necessarilly straight); therefore we have a cylindrical-like density profile with its 
associated cylindrical-like potential. 
Following this principle, and at the scales we are interested in in this paper (filaments between high mass haloes), 
we will assume a filament is of higher quality than another one if it is more likely to satisfy the previous conditions.
A stronger cylindrical-like density profile (indicated by the DM density between consecutive halo pairs
in the skeleton) above the background will produce a stronger collapse of matter towards the skeleton, and
smaller gaps between filament backbone members will better guarantee the continuity of the filament.
The complex geometries and different scales characterising filaments, along with the facts that there is no known
density profile a filament should follow and that they are unstable structures, make it difficult to set the 
values for these two parameters that will ensure a high quality sample of
filaments.  Instead we simply assume that a higher characteristic density and smaller gap parameters imply a
higher quality filament.
 }

\begin{figure*} 
\centering 
\vspace{0.1in}
\includegraphics[height=1.2\linewidth,angle=0]{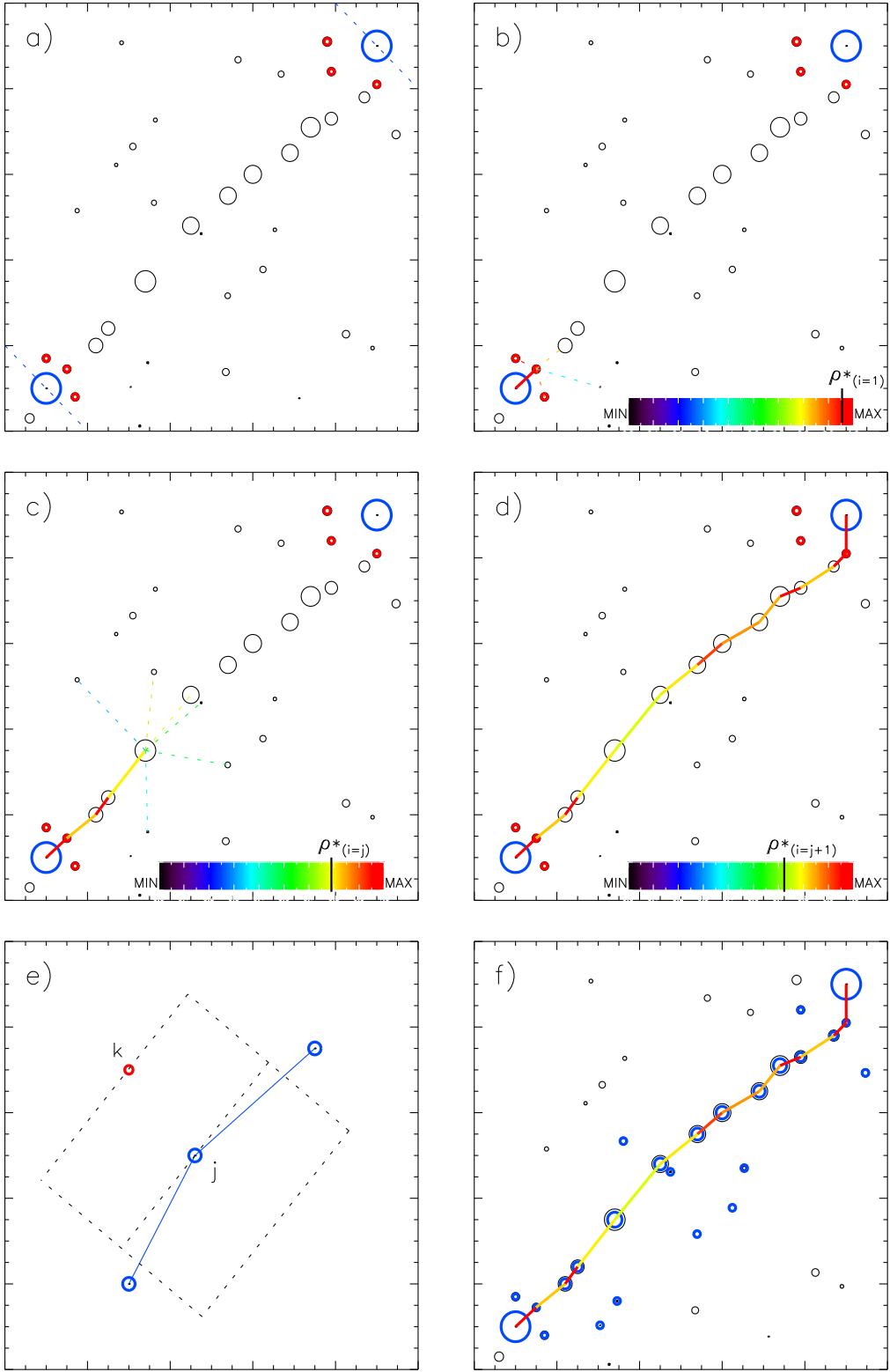}
\caption{
\label{figuremet} Filament detection method steps. Details of each step in text.}
\end{figure*}

\begin{figure*} 
\centering 
\vspace{0.1in}
\includegraphics[height=.7\linewidth,angle=0]{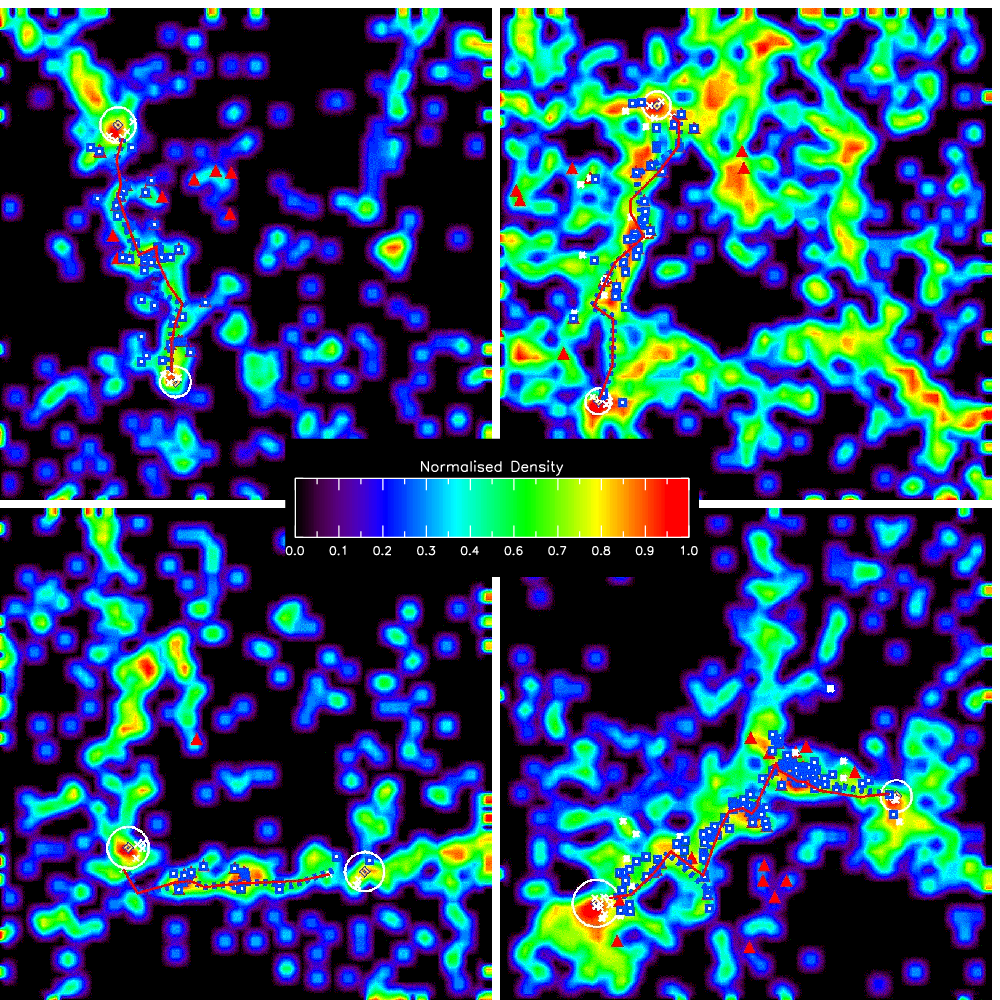}
\caption{
\label{figure3} Four examples of detected filaments. The red solid lines show filament skeletons, the blue dashed lines show the re-centred skeleton.   The white asterisks correspond to haloes at distances from the filament $r<r_0$, whereas blue squares show haloes at distances $r<r_1$.  The red triangles show haloes with $E_P<0$.}
\end{figure*}

\subsection{Description of the algorithm}

 {Figure \ref{figuremet} shows a cartoon depiction of some of the steps followed by the algorithm to identify
filaments for a particular node pair; in the figure, circles represent halo positions and their virial radii.}
We identify filaments in the following way,
\begin{enumerate}
\item We select a node tracer pair  {(indicated by blue circles in the figure)}.
\item 
We follow the segments of highest local DM density given by the characteristic density $\rho*$. This defines the filament backbone or skeleton.
For this we define a set of threshold  densities $\rho_{th}(i)$ with $i=1..N$, in the range set by the minimum and maximum densities in the full density field.
\item For each node we generate a list of neighbour haloes just outside the virial radius in the half hemisphere that points 
to the other node. These neighbours will be labeled as start haloes associated to the node from which we will start 
the filament search. End haloes will be the neighbours associated with the other node in 
the node pair in the half hemisphere pointing back to the start node. 
 {In panel a) of Figure \ref{figuremet}, the blue dotted lines indicate the half hemispheres of the nodes that point to the other node; red circles mark the haloes at the start and end nodes.}
\item {  The first attempt at identifying a filament is done starting at the highest density threshold $\rho_{th}(i=1)$.}
\item The process is iterative selecting the start halo with the highest local DM density with respect to the start node,
characterised by a local density greater than $\rho_{th}(i)$.  
A halo that satisfies this condition becomes part of a possible skeleton, and we search for neighbours of this
new skeleton member using the same conditions.
If there are no new neighbours satisfying this, we go back to the previous halo from where we will choose a different neighbour to restart the procedure.
 {Panel b) depicts this step.  The colours of the lines (solid and dashed) connecting pairs of haloes 
correspond to the local characteristic DM density (densities are shown in the colour-scale bar at the bottom of the panels).
As can be seen, we start with the maximum characteristic density threshold $\rho_{th}(i=1)$ denoted by a vertical black 
line in the colour bar. We choose the start halo (the one connecting with the start node located near the bottom of the panel) which has
four neighbour candidates (connected by dashed lines to the start halo) for skeleton members, but 
two neighbouts are neglected since they are also start haloes.  This leaves two remaining candidates, but none of them 
are characterised by densities higher than the threshold, and we are not able to find a filament at this density 
threshold. }
\item We repeat the last step {  with a different start halo} until any of the end haloes of the other node are reached, or until there are no more haloes satisfying these rules.
\item If no connection to the other node is found, we move down to the next {  lower} density threshold step $\rho_{th}(i+1)$, and go back to step v. 
 {Panel c) shows the skeleton after lowering several times the density threshold down to the point 
where the skeleton contains four members (connected by the solid lines). 
However, the fourth skeleton member has no neighbour candidates (connected by dashed lines to the fourth member) with
characteristic DM density greater than the current threshold.}
\item We will always find a set of connected points (a filament backbone) between two nodes for a sufficiently low value 
of $\rho_{th}$ density.  Higher values of this density imply stronger filament contrasts.
 {Panel d), shows the result when a first skeleton was completed between the two nodes, for
a sufficiently low density threshold.}
\item We re-centre the local centre of mass of the filament skeleton using its immediate Voronoi neighbours.
\end{enumerate}

Having a well defined backbone, we start adding skeleton neighbours to the filament and computing filament characteristics, in the following way,

\begin{enumerate}
\item For any given halo $k$ we find the nearest skeleton member $j$ { (shown in panel e)}

\item We measure the mass contained in a cylinder around the skeleton at the position of the skeleton halo $j$.  The cylinder height is $H=(D_{j,j+1}+D_{j,j-1})/2$ 
and its radius $R=D_{k,j}$.
 Using this mass and the difference between the average velocities of the haloes within that cylinder and that of halo $k$, projected in the plane perpendicular to the cylinder, we compute the total halo energy in the plane, $E_P$.
 {In panel e) of Figure \ref{figuremet}, the cylinder is depicted by black dashed lines. The cylinder axis(middle black dashed line) is tangent to the filament at halo $j$ as inferred using the two immediate neighbour skeleton members.}

\item We compute the orbit time $t$ around the cylinder for halo $k$ assuming that the distance $D_{kj}$ is the semi-major axis of the orbit.  This timescale only uses information on the potential energy and does not require peculiar velocity data.

\item We select all haloes with $E_P < 0$ and calculate their median
orbit time $t_1$; we define $r_1$ as the radius containing $80\%$ of these haloes.  This sample can only be obtained from haloes with peculiar velocity information.

\item We select all haloes with $E_P < 0$ and $t \leq t_F$, with $t_F$ a fixed input parameter, and we define $r_0$ as the radius where $80\%$ of these haloes are contained.  This defines a sample using $E_P$ measurements and it therefore needs peculiar velocity information to be constructed.

\item We select all haloes with $t \leq t_F$, and define $r_2$ as the radius where $80\%$ of these haloes are contained.  
This selection can be done with position and mass information alone and  {does not require dynamical information.}

\item Finally, we also select all haloes with $t \leq t_1$, and we define $r_3$ as the radius containing $80\%$ of these haloes.
This selection also requires velocity information and is used to assess the importance of the binding energy 
condition against that of the orbital timescales.
\end{enumerate}

All haloes closer to the skeleton than $r_1$ will be selected
as filament members in the simulation. 
 {Panel f) of Figure \ref{figuremet} shows the resulting filament, where blue circles correspond to
haloes belonging to the new filament; the remaining nearby haloes are too far away from the filament and do not satisfy 
the membership conditions. }

\section{Results}

Figure \ref{figure3} shows four detected filaments in the simulation, where the halo density projected onto the $x-y$ plane is shown
in a colour scale, the
skeleton is shown as red lines, and the re-centred skeleton as blue dashed lines. The nodes are indicated by circles with radii equal to the halo virial radius.  White points denote all haloes {lying closer than $r_0$ from the filament skeleton, and blue boxes denote haloes closer than $r_1$}.  The red triangles are for haloes with $E_P < 0$. 
All the filaments contain segments with only either a few or no bound haloes, at least according to our definition.  

We bear in mind the possibility of undetected bound haloes since in our energy calculation we do not take into account nearby structures other than the filament.
In order to produce a more precise energy calculation one would need to use velocities from other sections of the skeleton instead of only from the nearest skeleton section;  filaments show a very complex velocity structure where nodes sometimes move towards each other (they may merge in the future) or away from each other, making filaments suffer stretching, elongations, torsions, and even rotations. 
However, the incompleteness in the sample of bound haloes
should not affect our estimate of the mean effective radius of the filament ($r_1$) which we use to define filament membership.

In the upper-left and bottom-left panels of Figure \ref{figure3} the filaments show excellent density contrasts, but also show a gap (near the top node in the upper-left panel, and near the left node in the bottom-left panel).  This shows the importance of adopting a gap parameter that allows the existence of these features in selected filaments to some degree.  The filaments in the right panels are of higher quality than those on the left since they do not show important gaps.   The section of the filament on the upper-right panel seems not to follow the highest density path due to projection effects (the filament follows a path that enters the page, along the z-axis).

\subsection{Filament properties}
We apply the method to the numerical simulation described in Section 2, using 
a minimum skeleton characteristic density $\rho*_{min}=3\rho_{mean}$ and no gap restriction, limiting the node pairs 
to relative distances lower than $65$Mpc$/$h.
 
\begin{figure} 
\includegraphics[width=0.9\columnwidth,angle=0]{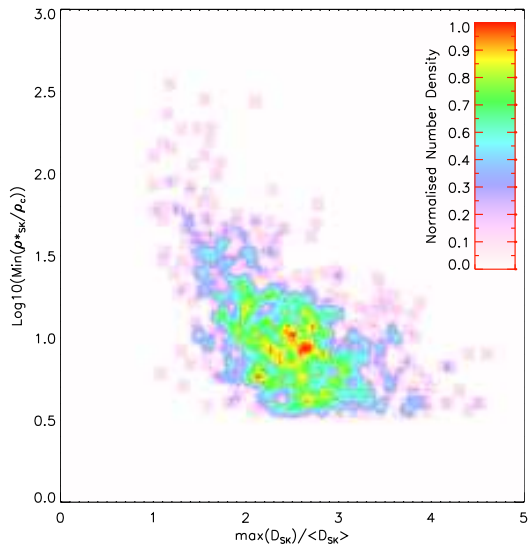}
\caption{
\label{figure4} Filament quality parameters.
Minimum skeleton density ($\rho*_{min}$) as a function of
the Gap size ($max(D_{SK})/<D_{SK}>$).
}
\end{figure}

{Out of the $3385$ node pairs, $1326$ are successfully connected via filaments; we will refer to this
first identification as the full sample.  We select an additional subsample of $467$ filaments which satisfy 
the additional conditions of $\rho*_{min}$ above the median of the full sample, and $max(D_{SK})/<D_{SK}>$ below the median; this sample is termed the high-quality subsample and contains $33\%$ of the filaments in the full sample.  The
separation between backbone members in the full and high-quality samples are, on average, $4.80\pm0.03$ 
and $4.19\pm0.03$ times the sum of their virial radii, respectively}.  As was
mentioned above, all the detected filaments connect nodes separated by at least the sum of their virial radii.
Figure \ref{figure4} shows the relation between  gap and density parameters for the detected filaments which show clear trends of larger gaps at lower densities.

\begin{figure*} 
\centering 
\includegraphics[height=.45\linewidth,angle=0]{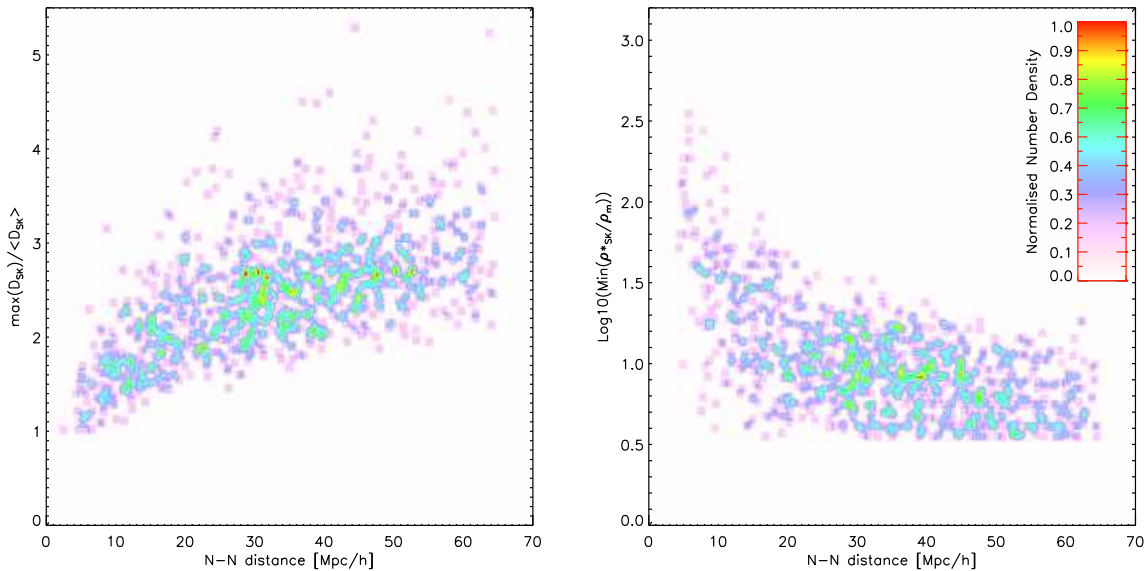}
\caption{
\label{figurex} Gap (left) and density (right) quality parameters as a function of node separation.}
\end{figure*}

Figure \ref{figurex} shows the dependence of the quality parameters on node separation for the full sample. There are clear correlations, {particularly for filaments shorter than $20$Mpc$/$h, which suggests that shorter node separations produce higher quality filaments.}

{
When studying the properties of the filaments detected using our automated procedure, it will be useful to compare
with a previous detection.  In particular, we will use the results from \citet{colb} 
who detected $228$ filament in a DM simulation by eye
using the smoothed DM density distribution. This filament sample can not be compared directly with our 
results, since the selection criteria are very different. However, both samples are the result of
restricting the search to filaments connecting neighbouring haloes above $10^{14}$h$^{-1}M_{\odot}$.
The main differences between the two samples arise from, i) Colberg et al. use the distribution of DM particles 
whereas we use halo positions, ii) they look for filaments using the $12$ nearest haloes inside cylinders 
of $7.5$h$^{-1}$Mpc of radius aligned along the node-node axis; in our case we look at all possible neighbour node pairs 
given by the voronoi tessellation with no volume constrain, iii) Colberg et al. define a true detection 
based on a visual criterion instead of using quality parameters, iv) they discard node pair connections when
other clusters lie inside the innermost $5$h$^{-1}$Mpc from the node-node axis, and we discard node pair connections 
when another cluster is closer than $2$ times its virial radius to the filament skeleton, v) they divide their 
sample in straight, off-centre and warped filaments.
Therefore, the reader must bear in mind that comparisons between these two samples, are not intended to validate 
any of the two samples, but to find general filament properties which are less sensitive to different
selection criteria.}

\begin{figure} 
\centering 
\includegraphics[width=.95\columnwidth,angle=0]{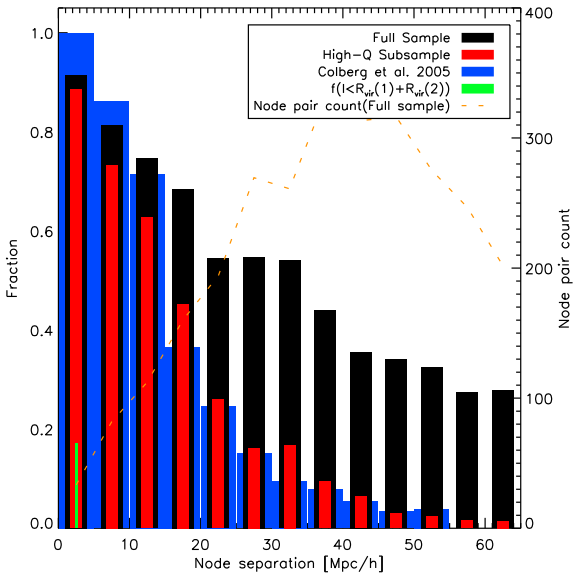}
\caption{
\label{figurec1} Fraction of node pairs with detected filaments as function of node pair separation (left y-axis). The green bar on the first bin shows the fraction of node pair connections separated by less than the sum of their virial radii. The total node pair count as a function of node separation for the full sample is shown as an orange dashed line (its scale is indicated on the right y-axis).  Black bars show the same fraction for the full sample of quasars, red bars are for
the high quality subsample, and the blue bars are for the Colberg et al. sample detected by eye.}
\end{figure}

{The node pair connections are given by the voronoi tessellation method, which instead of
selecting the $n$ nearest neighbours, chooses neighbours such that the line that connects the pair passes only 
through the voronoi cell around each node. This ensures that any point along the segment is nearest to one 
of the two nodes and not to other haloes. The node pair count of the full sample as function of the node 
separation is shown in Figure \ref{figurec1} as an orange dashed line (the scale of the counts in $5$h$^{-1}$Mpc bins  
is given by the right y-axis).   The number of pairs grows almost linearly with the separation almost up to 
$40$h$^{-1}$Mpc, and then it decreases for larger node separations.  In addition, Figure \ref{figurec1} shows 
the fraction of node pairs with detected filaments as a function of node separation 
(left y-axis scale).  The full sample (solid black bars) is characterised by a decreasing fraction of connected pairs
via filaments as the separation increases; this fraction is nearly $90\%$ for separations shorter than $5$h$^{-1}$Mpc, and at the largest separations the fraction is reduced to $30\%$. In the case of the high quality subsample (red bars) the abundance of filaments decreases much faster with fractions below $25\%$ for nodes separated by more than $20$h$^{-1}$Mpc.} 

{Figure \ref{figurec1} also shows the fractional abundance obtained by \citet{colb}.  Even though their 
selection procedure is different from ours, the resulting dependence of this fraction with pair
separation is similar to our results for the high quality subsample.  
Our method does not consider haloes inside the virial radii of nodes, which means that we do not detect most of the 
filamentary structure connecting two nodes separated by distances shorter than the sum of their virial radii. 
The green bar shown for separations  shorter than $5$h$^{-1}$Mpc in Figure \ref{figurec1} indicates the fraction 
of node pairs whose separation is shorter than the sum of their virial radii for this range of separations. 
Most of the pairs represented by the green bar should be connected by filaments \citep{pimbblet}, 
since these are overlapping bound systems which share matter \citep[i.e.][]{die,tit}; furthermore, this behavior should extend up to node separations of a few virial 
radii (approximately three times the virial radius), 
which can be associated to the infall region of haloes \citep{dia,piv}. 
Taking into account the mass resolution of our numerical simulation,
the mass in such bridges is mostly in the form of a smooth DM particle distribution, with only a few subhaloes 
aligned within the bridge. This makes it more difficult to detect them with our method even if we also used subhalo
positions; as a consequence we have chosen not to include them in the search.   
A possible way to overcome this would be to use the DM particle distribution, or to run re-simulations of 
these regions with higher resolution, enough to resolve several subhaloes per node. 
The result of such a study would likely change our fraction of detected filaments for node separations below $5$h$^{-1}$Mpc, 
which we are underestimating at present; in the case of \citet{colb}, they find that most of the halo pairs
within this range of separations are connected via filaments. }

In the case of filaments detected in the 2dFGRS, \citet{pimbblet} find a fractional abundance of filaments similar 
to our full sample results; however, their selection criteria are also different from the one we have applied
to the simulation.  In particular, they also identify filaments by eye and use galaxy positions; therefore
in order to make an appropriate comparison it would be necessary to apply our method to realistic 2dFGRS mock
catalogues, or directly on the 2dFGRS catalogue.

\begin{figure*} 
\centering 
\includegraphics[height=.85\linewidth,angle=0]{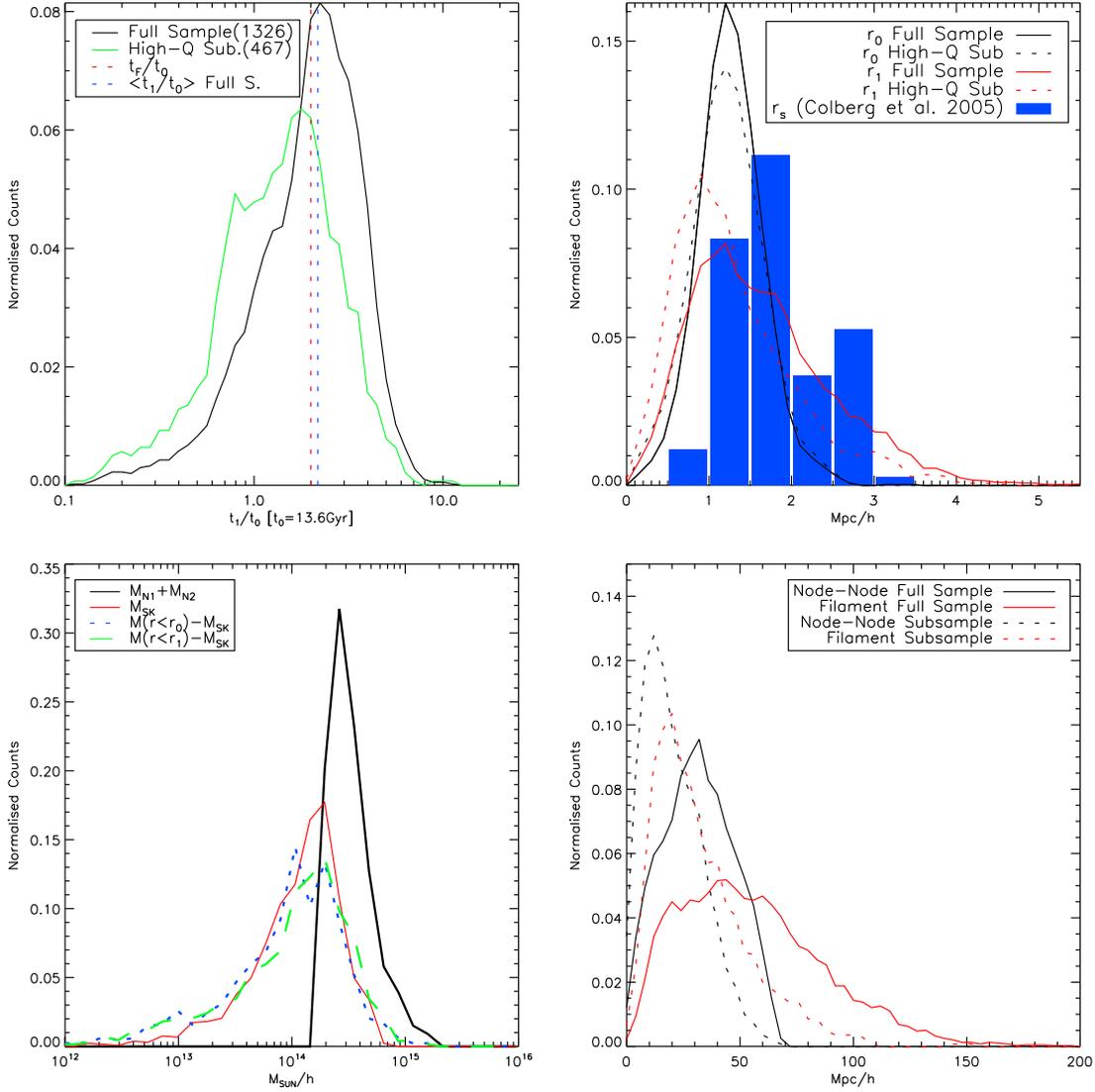}
\caption{
\label{figure5} Distribution functions of properties of the detected filaments.  The samples of filaments
detected using our automated method are shown in different line types (explained in the key).  The statistics from the
\citet{colb} filament sample are shown as barred histograms.  }
\end{figure*}

The main properties of the detected filaments are shown in Figure \ref{figure5}. The top-left panel shows 
the distributions of $t_1$ (the median orbit time for haloes with $E_P<0$), where it can be seen that the high
quality filaments are characterised by lower orbit times as expected since these filaments have higher density contrasts
and are more concentrated than the full sample.  
The samples shown in the figure are obtained by setting
$t_F=2t_0$ (vertical red dashed line) which is slightly lower than the median of $t_1$ (indicated by 
the vertical blue dashed line).  This latter value can be used when detecting filaments  {without dynamical} data
since the orbit time distributions shown here are relatively narrow (most of the filaments show similar orbital 
timescales).

The top-right panel of the figure shows the distributions of the parameters $r_0$ and $r_1$ (line types 
are indicated in the figure key) described in the 
previous section.  As can be seen, a fixed orbit time produces a narrow distribution of $r_0$ but a wider
distribution of $r_1$ which is obtained using $t_1$.  However, the peaks of both distributions are located at
approximately $1.3$Mpc$/$h.  It is also noticeable a very slight shift towards smaller radii for the high 
quality subsample in both cases, {an effect which is stronger for the $r_1$ parameter, indicating a dependence 
of the $t_f$ value with the quality of the filaments}. Therefore, better quality filaments seem to be more 
concentrated while preserving similar thicknesses with respect to the filaments in the full sample. {In addition, 
the figure also shows
the scale radius $r_s$ computed by \citet{colb} for their sample of filaments (blue bars).  In their notation
$r_s$ defines the radius where the density profiles of straight filaments starts to follow a $r^{-2}$ relation. 
Our definition of $r_1$ indicates a scale radius containing $80\%$ of the bound haloes with orbit times below the median.
Even though both definitions are conceptually different, they account for the scale radius where 
$\approx 50-80\%$ of the filament mass is contained.  In general, for a given filament, $r_s$ is a more precise 
computation of the edge of the filament, but requires the DM particle distribution to be calculated; $r_1$ is easier 
to compute since it only requires halo positions; however, it can underestimate the filament edges depending on 
the density profile and density contrast.  Therefore, despite the fact that the comparison is made among two
quantities with different definitions, as well as different filament samples, it is interesting to
note that the distributions of $r_1$ and $r_s$ show similarities; the latter only shows a slight shift 
towards larger radii.
As can be seen, the characteristic radius which defines a filament shows a narrow distribution
with preferred values of $1$ to $2$h$^{-1}$Mpc, even when using filaments of different quality or using a sample of 
filaments selected by eye.  In all cases, however, the lengths of the filaments are similar and are traced 
by halo nodes with masses above $10^{14}$h$^{-1}M_{\odot}$.  }

The Bottom-left panel of Figure \ref{figure5} shows the distribution of mass for different filament components (line
types are show in the figure key);
all the distributions are shown for the full sample of filaments.  As can be seen, this tracer node mass selection
produces skeletons and filament envelopes less massive than the filament nodes.  Both of these two
components show similar distributions, with differences only at the low-mass-end. 
Notice that when using either $r_0$ or $r_1$ the resulting filament mass is practically the same. 
This shows that the detection of filaments using a fixed orbit time ( {when no dynamical information
is available}) will provide
reliable filament mass measurements. 
In the case of the high quality filaments, we find that the masses of the skeleton and the surrounding filament
shells are lower than for the complete filament sample, since the former are shorter in length (as can be seen 
in the bottom-right panel of the figure). We find no clear dependence of filament mass on their node masses.

The bottom-right panel of the figure shows the distributions of node pair separation and of filament extension (line
types are indicated in the figure). 
The filament extension is obtained by adding 
the distances between consecutive filament member 
 positions (i.e. in a discrete number of segments) along the filament.  
The node separation is on average
smaller than the filament length, which indicates that most filaments are warped. 
The distribution of node pair separation peaks at $\approx 32$Mpc$/$h for the full sample, and at $\approx 15$Mpc$/$h for 
the high quality subsample.  The filament lengths also show a peak at shorter values for the high quality
subsample. {When analysing the ratio between these two quantities in both, the full and high quality samples,
it can be seen that regardless of quality, longer filaments are more warped than shorter filaments; i.e. in the full sample, filaments with 
node separations below $30$h$^{-1}$Mpc are on average $13\%$ larger than their node separation; this value increases 
to $40\%$ for larger node separations.}
{The filaments studied with Shapefinders in the Las Campanas Redshift Survey \citep{bha} are
characterised by lengths of $50$h$^{-1}$Mpc to $80$h$^{-1}$Mpc. In this work, we have found shorter high quality 
filaments but we have also required node pair separations $<65$h$^{-1}$Mpc. 
It should also be borne in mind that 
in most cases these filaments are only segments of considerably longer structures with more than two nodes (shapefinders
are insensitive to the number of nodes in a filament).
}

\begin{figure} 
\centering 
\includegraphics[width=.95\columnwidth,angle=0]{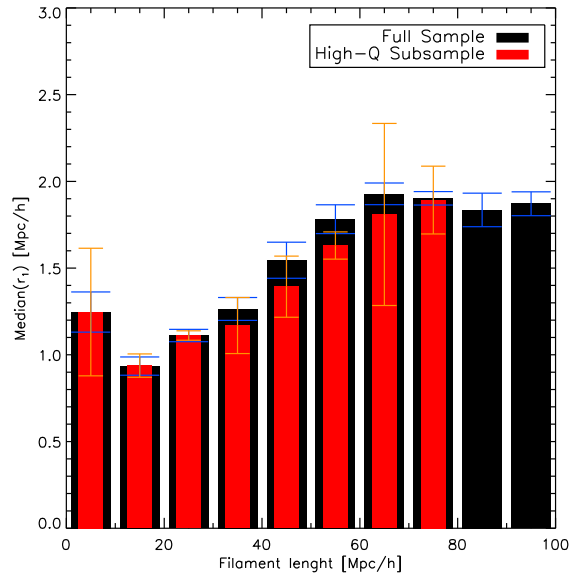}
\caption{
\label{figure6} Filament thickness (as measured by $r_1$) as a function of filament length, for the full sample (black) and high quality subsample (red).}
\end{figure}

Figure \ref{figure6} shows the relation between filament thickness ($r_1$) and filament length.
{The error bars correspond to the standard deviation in the measurement of the median of $r_1$, and are 
computed using the jackknife method.  In the high quality subsample, we do not include filaments longer than 
$80$h$^{-1}$Mpc due to low filament counts ($<10$). }
As can be seen, there is a trend of thicker filaments for longer filament lengths in both samples (full
and high quality).
For the high quality sample, the median value of $r_1$ for filaments with lengths between 
$0$ and $10$h$^{-1}$Mpc is $1.11\pm0.19$h$^{-1}$Mpc,
and for lengths between $60$ and $70$h$^{-1}$Mpc it is $2.01\pm0.29$h$^{-1}$Mpc
(a significance of more than $3\sigma$ for a difference between the longest and shortest filament lengths).  
This dependence can be a consequence of any or several of the following effects,
 i) all filaments feed their node haloes and shorter, less massive filaments will exhaust their mass first due to 
the higher infall velocity and node halo influence over a larger percentage of the filament length  (the influence
can extend out to several virial radii, \citealt{dia}), 
ii) shorter filaments are straighter than longer ones; therefore, in longer, warped filaments concave zones along 
the skeleton could attract haloes from larger distances, an effect that would be absent in straight-line filaments. The detailed
study of this possibility is beyond the scope of this paper and
will be treated in a forthcoming paper on filament shapes and environments.
iii) A higher probability to spuriously assign bound haloes at larger distances from the skeleton for 
longer filaments, but this is less likely since this effect is also present when using $r_0$ (which does 
not depend on a computation of energy) as a thickness indicator.

\begin{figure}
\centering 
\includegraphics[width=.95\columnwidth,angle=0]{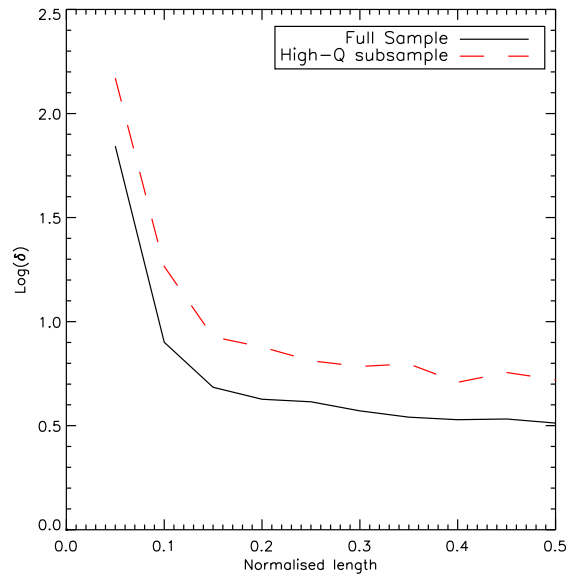}
\caption{
\label{figurelen} Average longitudinal filament over-density profile obtained using the interpolated voronoi 
density along the skeleton, as function of the normalised node pair separation.  We only show half of the
filament length since the profiles are symmetrical, on average.}
\end{figure}

{We study the variation of the mass density along the filament skeletons.  Figure \ref{figurelen} shows 
the average over-density as a function of the normalised node pair separation.  It should be borne
in mind that as we use the interpolated voronoi density 
obtained from the halo positions and their viral masses, the density only includes a fraction of the total matter 
(DM particles beyond the virial radii of haloes are not included in this estimate).  We exclude filaments with 
skeletons containing less than $6$ haloes, and the figure only shows half of the filament length since the profiles 
are symmetrical (on average). 
The figure shows a similar density profile to those found by \citet{colb}, where the over-density rises towards 
node centres, indicating that on average the infall regions of filaments extent up to $20\%$ of the filament length.
At larger distances from the nodes, the overdensity remains at nearly
constant values of a few times the average density.
The high quality subsample shows a similar profile although with higher density contrasts than the full sample. }

\begin{figure}
\centering 
\includegraphics[width=.95\columnwidth,angle=0]{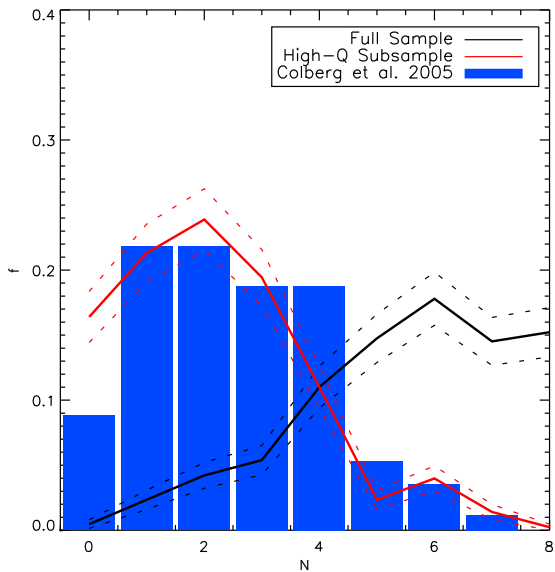}
\caption{
\label{figurenp} Fraction of nodes connected to $N$ filaments for the full sample of filaments (solid black lines), and 
for the high quality subsample (solid red lines).  In both cases, Poisson errors are shown by the dashed lines. 
The results from Colberg et al. (2005), are indicated as blue bars. }
\end{figure}

{We now study the number of filaments connected to individual nodes, and how this depends on the node properties.
Figure \ref{figurenp} shows the fraction of filaments connected to $0,1,2,...$ filaments for the full sample 
(black solid lines), the high quality subsample (red solid lines), and the \citet{colb} results (blue bars). 
The Poisson error amplitudes are shown as dashed lines for the full and high quality samples.  
In the full sample, most nodes are connected to $4-6$ filaments, indicating that allowing in all the detected 
filaments without applying any quality constraints does not provide realistic results, bearing in mind
the observational (Pimblett et al., 2005) and numerical simulation (Colberg et al., 2008) results on this statistics.
A better agreement with these estimates is obtained when using the high-quality subsample, in which case
most nodes are connected to $2$ filaments (and the distribution is very similar to that from the \citealt{colb} filaments). 
In general the number of filaments per node is strongly dependent on the quality of the filaments considered;
similar quality thresholds are needed in order to make meaningful comparisons.  
     }   
 {Given that the number density of filaments is three times higher for the full sample than for
the high-quality sample (simply due to the total number of objects in each sample) it can be expected that
the distribution of filament connections per node will also be a factor of three higher for the
full sample, that is $\simeq 6$ compared to $\simeq2$ connections for the full and high-quality samples, respectively
(as $\simeq83\%$ of the nodes of the full sample of filaments are connected by
high-quality filaments).}

\begin{figure}
\centering 
\includegraphics[width=.95\columnwidth,angle=0]{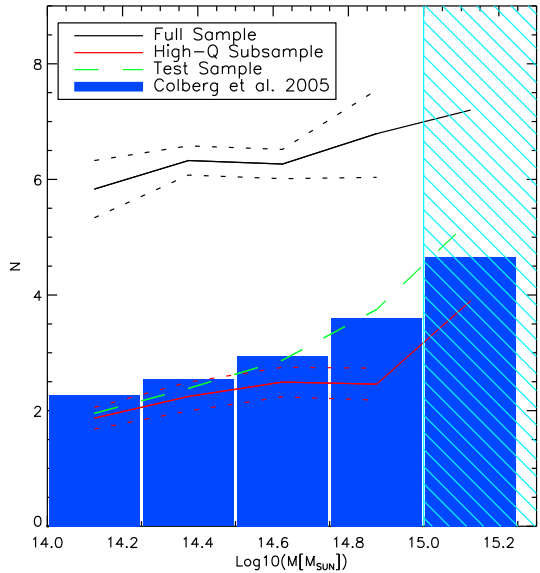}
\caption{
\label{figurenm} Number of filament connections per node as function of node mass.
Different line types correspond to samples selected in this paper (identified in the figure key);
the barred histogram corresponds to the sample of filaments in \citet{colb}.  The hatched area
shows the range of masses containing only $10$ node pairs in our numerical simulation.}
\end{figure}

{Figure \ref{figurenm} shows the average number of filaments per node as a function of node mass. 
In all cases this number increases with the node mass. Errors, shown as dashed lines for the full and high
quality samples, are obtained using the jackknife method;  errors are not shown for the highest mass bin 
$M>10^{15}M_{\odot}$ (cyan hatched region) due to the low number of nodes ($10$) at this end.
Nodes in the high quality subsample are connected to an average of $1.87\pm0.18$ filaments for the lowest
mass bin, a number that increases to $2.49\pm0.28$ for $M\simeq10^{14.9}M_{\odot}$; the significance of this trend
is higher than a $3\sigma$ level.
This behavior was also observed in the 2dFGRS by \citet{pimbblet}, and in a numerical simulation \citep{colb}, 
clearly indicating that more massive haloes are more likely to have a larger number of connected filaments.
This can be associated to the higher amplitude of clustering of more massive haloes characterising
random gaussian fluctuation fields in a
$\Lambda CDM$ cosmology \citep{pimbblet}.
}

{There is a number of possible issues that could affect this statistics that need to be borne in mind,
i) we do not use subhaloes, and therefore node pairs closer than
the sum of their virial radii could present filaments which we do not detect.  
Such close pairs will 
be more abundant for more massive haloes due to their higher local overdensities, therefore these undetected filaments 
could populate the high mass end of the figure \ref{figurenm}. 
ii) To avoid repeated filament segments, we discard filaments which are closer than $2r_{vir}$ to a third node, 
and \citet{colb} use a fixed value of $5$h$^{-1}$Mpc for a similar proximity condition. In both cases we 
could be missing short filaments in dense environments where nodes are more massive, have larger virial radii and 
are more strongly clustered; in such places this proximity constrain could be excessive. In order to
test this issue, we make a subsample of filaments applying the quality constraints used for the high quality subsample,
but allowing filaments closer to a third node when, a) the node pair separation is less than $10$h$^{-1}$Mpc, b) the
minimum density along the filament is greater than $10$ times the mean density, c) the sum of
the virial radii of the nodes is $ > 2.5$h$^{-1}$Mpc, 
d) the filaments are close to straight-line shapes.  These modifications, in conjunction with 
the intrinsic properties of voronoi tessellations for the node pair selection, ensures that it is very unlikely 
that the short filaments in this new sample are repeated segments of other detected filaments.  The reason
behind this is that for larger node pair separations, there will be larger distances from a node to node axis 
to a third node.  Otherwise the constrain of a common facet between node pair voronoi cells would not be fulfilled.
This test subsample is shown as green long dashed line in Figure \ref{figurenm}; as can be noticed the relation 
of filament connections as a function of mass becomes stronger.} 

\subsection{Application to observational data}

 {In the case of applying this method to galaxies, we can use luminosities instead of halo masses and detect 
filaments following the path of highest luminosity density.   In this case, as the light of a galaxy is more 
concentrated than the mass 
it is safer to assume that the voronoi density traces that of the luminosity in both, the high and low density regimes.}
 {In this case, the filament quality can be defined using a luminosity density parameter as well as a gap  parameter.}
However, it would become more difficult to measure a filament thickness since
 {in general there would be no information on the mass and, additionally,} 
there seldom is dynamical information to calculate binding energy conditions {  in galaxy samples}.  

 {A possible way to apply this method could use the skeleton brightness and a brightness threshold for filament 
membership where i) the distribution of filament thickness, and ii) the relation between filament thickness and
lenght match the results from a DM simulation where the settings on the quality parameters result in similar number 
densities of filaments.
These tests, and an application to observational data from the SDSS are part of a forthcoming paper.}

 {In the case of a sample with estimates of galaxy masses but no dynamical data, such as in nearby galaxies, 
it would be } 
possible to select filament members assuming that galaxies are bound 
to the filament, and requiring orbit times lower than $t_F$.  {In the simulation,} as can be seen in Figure \ref{figure5}, using a 
fixed orbit time allows to recover a distribution of $r_0$ (see Section 3.3 for the definitions
of $r_0$, $r_1$, $r_2$ and $r_3$) which, although slightly narrower, 
peaks at the same radius as when using the full energy calculation.  Also, the recovery of the
filament mass is only mildly affected by the use of $r_0$ or $r_1$ to select filament members.

\begin{figure*} 
\centering 
\includegraphics[height=.45\linewidth,angle=0]{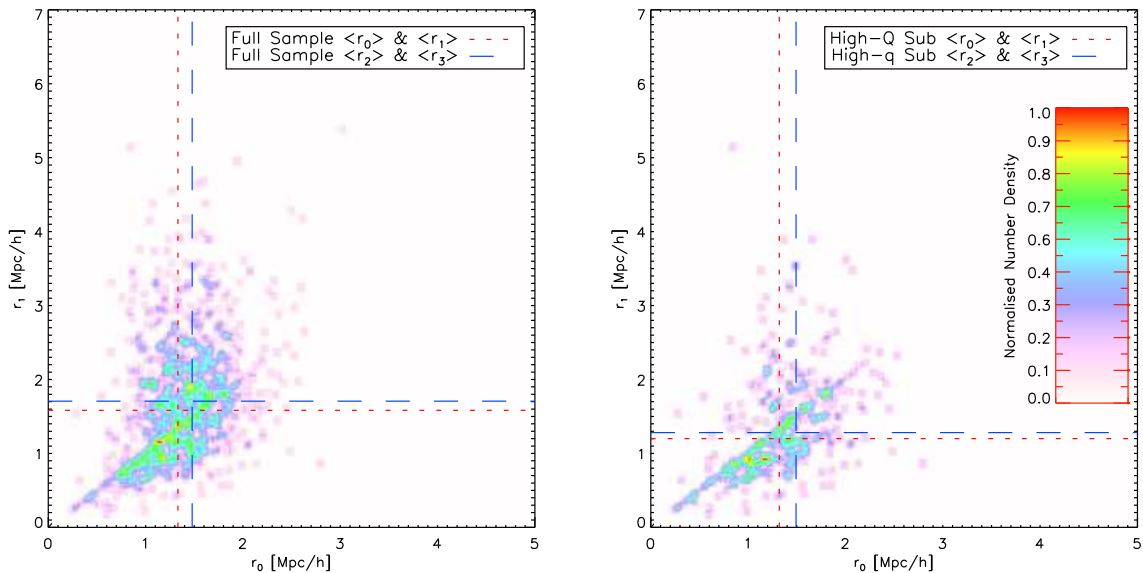}
\caption{
\label{figure7} $r_0$ vs. $r_1$ for the full sample of filaments (left panel) and for the high quality subsample (right).
The vertical and horizontal dotted lines show the median values of $r_0$ and $r_1$ (respectively), which
are quantities obtained using the full binding energy calculation.  The dashed lines show
the median values of $r_2$ and $r_3$ (vertical and horizontal lines, respectively), which are the equivalent to $r_0$ and $r_1$ for
the case with no dynamical information (and therefore no energy calculation).
}
\end{figure*}

Figure \ref{figure7} shows the relation between $r_0$ and $r_1$.  As can be seen, there is a linear relation between
these quantities for $r_1<median(r_0)$.  Filaments in the high quality subsample show a very
similar median $r_0$ and a slightly lower median $r_1$ than the full sample, an effect that probably
arises from the fact that filaments in the high quality subsample are shorter than in the full 
sample (see Fig. \ref{figure5}).  
In the case of the observational data with  {masses, but} no dynamical
information, the method would only provide measurements of $r_2$ which, when comparing the
vertical long dashed and dotted lines in both panels, can be seen to provide a good approximation
to $r_0$.  As the relation between $r_0$ and $r_1$ 
is reliable for thin filaments, $r_2<1.2$Mpc$/$h,
thick filaments will probably suffer from an under-estimation of their real thickness, particularly
if their quality is low.
Regarding $r_3$ (horizontal dotted lines), it can be seen that their median values are very similar to that of $r_1$, indicating that
if one can estimate the collapse time of bound objects to the filament, the membership obtained using this estimated time will provide
a good membership criterion.

\section{Conclusions}

We presented an automated method to detect filaments in cosmological simulations, using 
haloes above a fixed mass as tracers of filament nodes.
 {In addition, we proposed possible directions to improve this method to allow its use with observational data.}
As filaments cannot be treated as virialised structures as in the case of haloes, and as they are 
characterised by a wide range of lengths, it is a difficult task to identify them
automatically.  As a result these have been mostly identified by eye.
In this work we detect filaments using an automated algorithm that provides 
two filament quality parameters, i) a minimum skeleton characteristic 
density, and ii) a gap parameter given by the maximum distance between consecutive skeleton neighbours 
divided by the average consecutive skeleton neighbour distance in individual filaments.  A small gap parameter
and a high density parameter, ensure the best quality for a filament. The latter condition is equivalent
to request a high density contrast.

In our method we define the width of filaments using the median radius ($r_1$) that contains the haloes
gravitationally bound to the filament in the plane perpendicular to the filament skeleton, and that
are characterised by orbit or collapse times below an upper threshold. 
An application of the method  {to data without dynamical information} can be done since the radius $r_1$ shows a good
correlation with $r_0$ and $r_2$ ($r_1$ is obtained assuming that all the galaxies are bound to the filament 
and computing their orbit times based only on their positions and masses); the members are then selected
requiring orbit times below a fixed time $t_F$. The relation between $r_1$ and $r_2$ is one-to-one
for thin filaments below $r_0\approx 1.2$Mpc$/$h; in thicker filaments $r_2$ tends to slightly under-estimate the actual
width of a filament.

We have presented several filament properties which can be studied in observational catalogues such as the SDSS. 
In particular, a subsample comprising the $33\%$ highest quality filaments in our numerical simulations
shows very similar properties to  
filaments detected by eye in numerical simulations by \citet{colb},
\begin{itemize}
\item {Filament lengths 
are mostly concentrated below $50$h$^{-1}$Mpc, but
can extend to up to $150$h$^{-1}$Mpc}
\item {Shorter filaments are characterised by more straight-line geometries than longer filaments.  
Filaments with node separations 
below $30$h$^{-1}$Mpc are $13\%$ longer than the distance between their nodes;  this increases to $40\%$ for larger 
node separations. }
\item {The distribution of filament widths is relatively narrow and shows a clear peak at $d=3$h$^{-1}$Mpc. 
There are indications of an increase in the filament thickness 
as the filament length increases.}
\item {Nodes are connected on average to $2$ filaments, this number increases slightly with the node mass, 
reaching $\approx 3$ filaments per node for masses close to $10^{15}M_{\odot}$}
\item {
In the infall 
region around nodes
the average central skeleton density can be as high as a hundred times the mean density; 
at larger distances the density drops to a few times the mean density, and maintains
a roughly constant value along $20-80\%$ of the filament length.}
\item {There is a strong relation between length, quality, and straightness in the filament shape, where 
shorter filaments have better quality and are closer to straight-line geometries.}

\end{itemize}

 {
Similarities of the high-quality sample with the \citet{colb} results seem to indicate that
the natural by-eye criteria are strongly related to our quality parameters; a detection by 
eye selects high density contrasts and few gaps.  
We stress the fact that did not intend to match the properties of the \citet{colb} filaments, instead we simply
chose the mean values of minimum density and gap parameters to define our high-quality sample.  
}

The filament properties we have studied in this work are focused on the general characteristics
of filaments.  There remain many specific properties of filaments and of their galaxy
populations which can be related to several recent results such as
(i) the halo clustering dependence on the halo mass and on its formation time \citep{gao}, 
(ii) the correlations between halo concentration and spin with the local environment \citep{avr}, 
(iii) the fact that
galaxy spins are strongly aligned along filaments \citep{pim2}, (iv) 
the results using semi-analytic models obtained by \citet{ggg} which show several variations of 
galaxy properties with the local and large scale environment, as well as (v) other results showing
that galaxy formation should be strongly dependent on the large scale environment starting 
from their early stages of development, due for example to the delayed reionisation of filaments 
with respect to clusters as shown by hydro-simulations of the intracluster medium \citep{finl}.
A first step will be to compare observational galaxy properties in filaments, in particular their 
colours, star-formation rates and luminosities with results from semi-analytic models, to characterise 
some of the previously mentioned environment effects.

Several studies of galaxy properties in clusters and voids have opened the possibility to
expect important variations in the properties of haloes
or galaxies while embedded in filament-like environments, 
since the populations of galaxies and haloes are very different in
voids and clusters.
By converging to a standard filament classification and detection method, the study of galaxy properties
and halo assembly in filaments can be carried out with great detail to help understand
the reasons behind these important population changes.

\section*{Acknowledgments}
REG was supported by project ALMA-CONICYT 31070007 and
 MECESUP PUC0609 fellow. NDP acknowledges support from FONDECYT Regular 1071006.
This work was supported in part by the FONDAP Center for Astrophysics $15010003$ and by BASAL-CATA.

\end{document}